# Is the 2026 East Asia Summer Extreme Heat Forecast Credible?

Xiaolei Liu[1,2,3], Jingzhi Su[3,4*]

[1] College of Atmospheric Science, Nanjing University of Information Science and Technology, Nanjing 210044, China

[2] State Key Laboratory of Disaster Weather Science and Technology, CAMS, Beijing 100081, China

[3] Center for Earth System Modeling and Prediction of CMA (CEMC), Beijing 100081, China

[4] Key Laboratory of Earth System Modeling and Prediction China Meteorological Administration, Beijing 100081, China

* Corresponding author (email: sujz@cma.gov.cn)

**Abstract** The limited predictive skill of forecasts makes it difficult for decision-makers to act decisively. Advance assessment of real-time forecast credibility can strengthen decision-makers' resolve and confidence to act. Such an assessment can draw on real-time observations of large-scale background signals. This study evaluates how credible the 2026 East Asia summer temperature forecast is. Enhanced predictability of East Asia summer temperature can be indicated by the synergistic forcing of sea surface temperature anomalies (SSTAs) across three key oceanic regions: the tropical western Pacific, the Japan Sea-Kuroshio-Kuroshio Extension (K-KE), and the North Atlantic. Based on the latest observational data and model predictions, the SSTAs in these three regions maintain positive anomalies, which suggests that East Asia's summer temperature forecast skill will stay at a relatively high level in the coming summer. Based on the predictions, the following summer is expected to feature pronounced positive temperature anomalies over central and eastern China, the Korean Peninsula, and Japan, which may trigger regional droughts and place severe strain on power supply networks.

## 1. Introduction

Subseasonal to seasonal (S2S) forecasting, spanning lead times from two weeks to an entire season, represents a critical frontier in bridging the predictability gap between short-term weather forecasts and long-term climate projections (White et al., 2022; Vitart and Robertson, 2018). Summer surface air temperature (SAT) over East Asia exerts profound impacts on human health, energy consumption, and water resource management (Cohen et al., 2020; Johnson et al., 2018). However, constrained by the inherently chaotic nature of the atmosphere, current models still exhibit highly limited prediction skill at S2S timescales (Merryfield et al., 2020; Lorenz, 1969). Given the still-limited skill of climate prediction, decision-makers often struggle to translate forecast results into substantive action, even when confronted with significant predicted anomalies. Nevertheless, mounting evidence indicates that under specific oceanic and atmospheric background conditions, S2S forecast skill can intermittently achieve exceptionally high levels for extended periods, a phenomenon recognized as "forecast skill windows" for S2S prediction (Liu et al., 2024; Mariotti et al., 2020). A priori evaluation of the credibility of real-time forecasts may enhance decision-makers' determination and confidence in taking action, which can be informed by real-time observed large-scale background signals, e.g., sea surface temperature anomalies (SSTA) (White et al., 2017).

Current climate monitoring reveals rapid anomalous warming of sea surface temperatures in the equatorial central-eastern Pacific. Multiple institutions predict a high likelihood of a strong El Niño event in 2026 (Fig. 1).

Under El Niño conditions, anomalous responses of the tropical climate system exhibit relatively high stability (Alexander et al., 2002; Wang et al., 2000). In contrast, extratropical climate is subject to considerable uncertainty, even under identical El Niño forcing, due to the influence of mid-latitude variability and the intrinsic instability of the East Asian summer monsoon–ENSO relationship (Xie et al., 2016; Wang, 2002). Current predictions for East Asian climate exhibit marked diversity, which hinders decision-makers from making informed judgments and taking decisive action.

Building on this context, this study evaluates how credible the 2026 East Asia summer temperature forecast is, by leveraging the precursor SSTA signals. The configuration of large-scale SST anomalies influences East Asian climate by modulating mid-latitude atmospheric variability (Liu et al., 2026; Ding and Wang, 2005). The persistent stability of these large-scale SST anomalies provides an a priori criterion for assessing S2S forecast skill.

# 2. Methods and Data

## 2.1 Data

The observational data were sourced from the ERA-5 reanalysis dataset (Hersbach et al., 2020), which includes daily 2-meter air temperature (T2m) and SST data, with a spatial resolution of 1°×1°. Anomalies within the observational dataset were obtained by removing the daily climatology, which was calculated based on the data from 1991 to 2020. This method eliminates both the seasonal cycle and the seasonal mean when calculating the anomalies.

The model data utilized are the hindcast data from the ECMWF model within the S2S prediction project dataset (Vitart et al., 2017). The ECMWF model generates forecasts extending to a lead time of 46 days. Each forecast comprises 11 ensemble members (one control and ten perturbed members), with a spatial resolution of 1.5° × 1.5°. The model data climatology was calculated using the model's hindcast data for each version. The climatology of the forecast data was calculated using the 20 years of hindcast data. Anomalies were then obtained by subtracting this climatology from the forecast members for each version, thereby effectively removing the systematic seasonal signals.

El Niño events were identified using the Relative Oceanic Niño Index (RONI) from the NOAA Climate Prediction Center (CPC). The RONI is calculated as the 3-month running mean of ERSST.v5 SST anomalies in the Niño-3.4 region (5°N–5°S, 120°W–170°W) relative to a 1991–2020 base period (Huang et al., 2017). Following the CPC definition, an El Niño event is identified when the RONI is ≥ 0.5°C for at least five consecutive overlapping 3-month periods that encompass the target summer season (June, July, and August; JJA) (Trenberth, 1997; Kousky and Higgins, 2007).

## 2.2 Skill Measures

To quantitatively evaluate the prediction skill of subseasonal-to-seasonal SAT anomalies for historical summer hindcasts sharing the same precursor signals, the area-weighted pattern correlation coefficient (PCC) was employed as the primary metric. This correlation is calculated between the model-predicted and observed SAT anomalies strictly over the terrestrial regions of East Asia, utilizing the cosine of latitude as the weighting factor for all land grid points. The PCC measures the spatial consistency between the forecast and observed fields, with its value ranging from -1 to 1. A value closer to 1 indicates higher spatial similarity and better forecast skill. The specific formulas are as follows:

$$PCC = \frac{\sum_{i=1}^{M}\sum_{j=1}^{N}\{\omega_{i,j}[(x(i,j)-\bar{x})*(y(i,j)-\bar{y})]\}}{\sqrt{\sum_{i=1}^{M}\sum_{j=1}^{N}\{\omega_{i,j}[x(i,j)-\bar{x}]^2\}\ \sum_{i=1}^{M}\sum_{j=1}^{N}\{\omega_{i,j}[y(i,j)-\bar{y}]^2\}}}$$

In the formula, $x(i,j)$ and $y(i,j)$ denote the observed value and the forecast value, respectively, at the grid point located at the $i$-th longitude and the $j$-th latitude. $M$ and $N$ are the total numbers of grid points in the meridional and zonal directions, respectively, within the study area, $\bar{x}$ and $\bar{y}$ represent the spatial mean values of the observed and forecast fields, respectively.

## 3. Results

Recent climate monitoring data and predictions from leading international ensemble systems, including the North American Multi-Model Ensemble (NMME) and Copernicus Climate Change Service (C3S), predict persistent warm SSTA in the equatorial central-eastern Pacific in the following seasons of 2026, highly likely evolving into a strong El Niño event (Fig. 1).

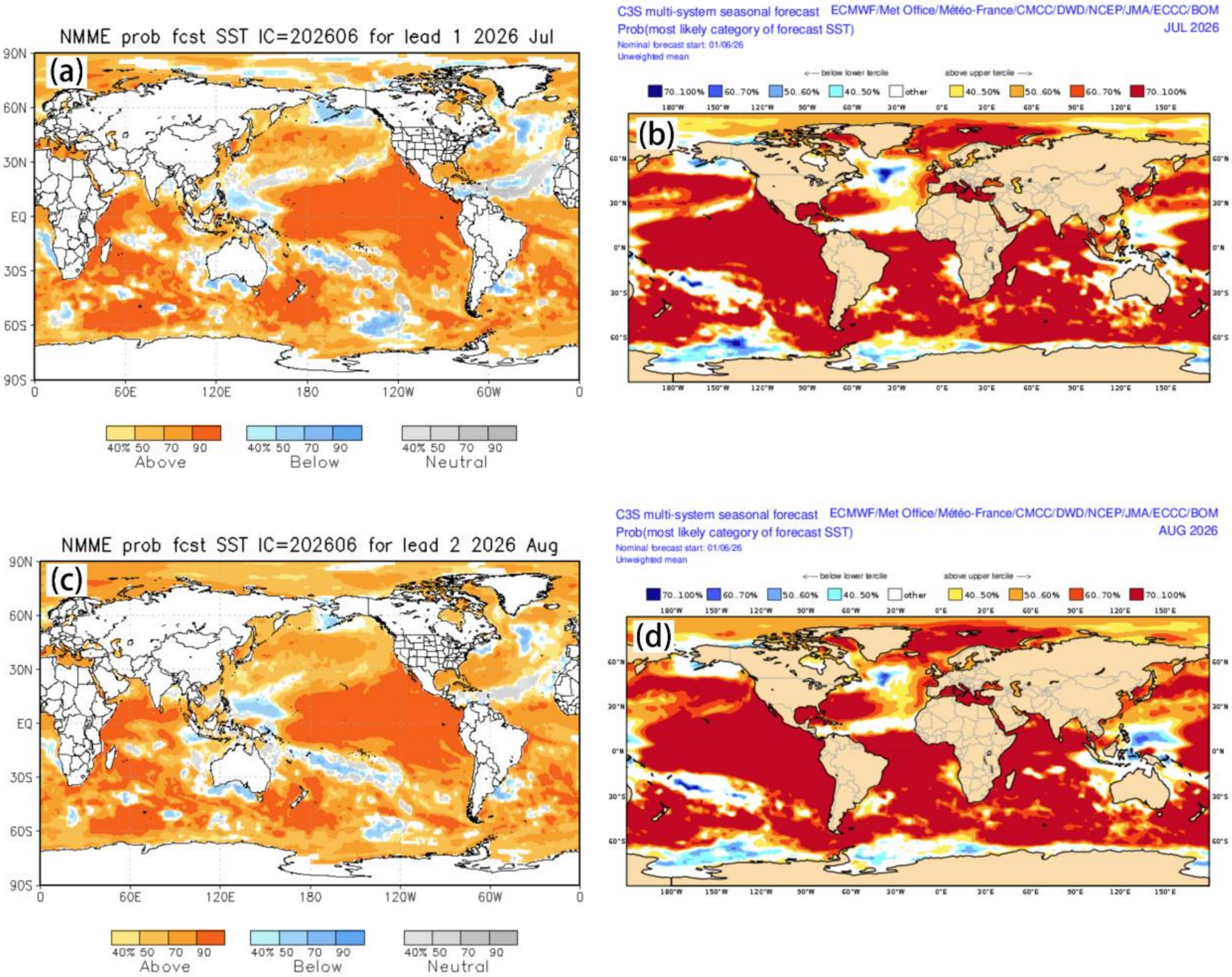

**Fig. 1.** Probabilistic forecasts of SSTAs from multi-model ensembles. The maps display the predicted probabilities of global SSTAs falling into anomalously warm (above-normal), anomalously cold (below-normal), or near-normal states, as generated by two multi-model systems. Panels a and c represent the forecasts from the NMME for July (Lead-1) and August (Lead-2) of 2026, respectively. Panels b and d show the corresponding forecasts from the C3S ensemble system for July and August 2026.

Typically, El Niño triggers widespread climate anomalies across the globe via atmospheric teleconnections. However, the climate over the East Asian monsoon region is modulated not only by tropical Pacific SSTs but also by the synergistic influences of multiple factors, including mid-to-high-latitude atmospheric circulations, which are potentially modulated by SSTA in mid-latitude oceans and land surface processes. Consequently, the specific impacts of El Niño exhibit substantial case-by-case variability. This can be clearly supported by statistical analyses of historical observations. During the developing phase of past El Niño summers (July–August), the spatial patterns of temperature anomalies across East Asia have varied, exhibiting inconsistency (Fig. 2). This robustly demonstrates that relying solely on El Niño as a single precursor to predict summer temperatures over East Asia inherently entails substantial uncertainty. A comprehensive evaluation of multiple factors—including mid-latitude SST anomalies—is essential for East Asian climate prediction.

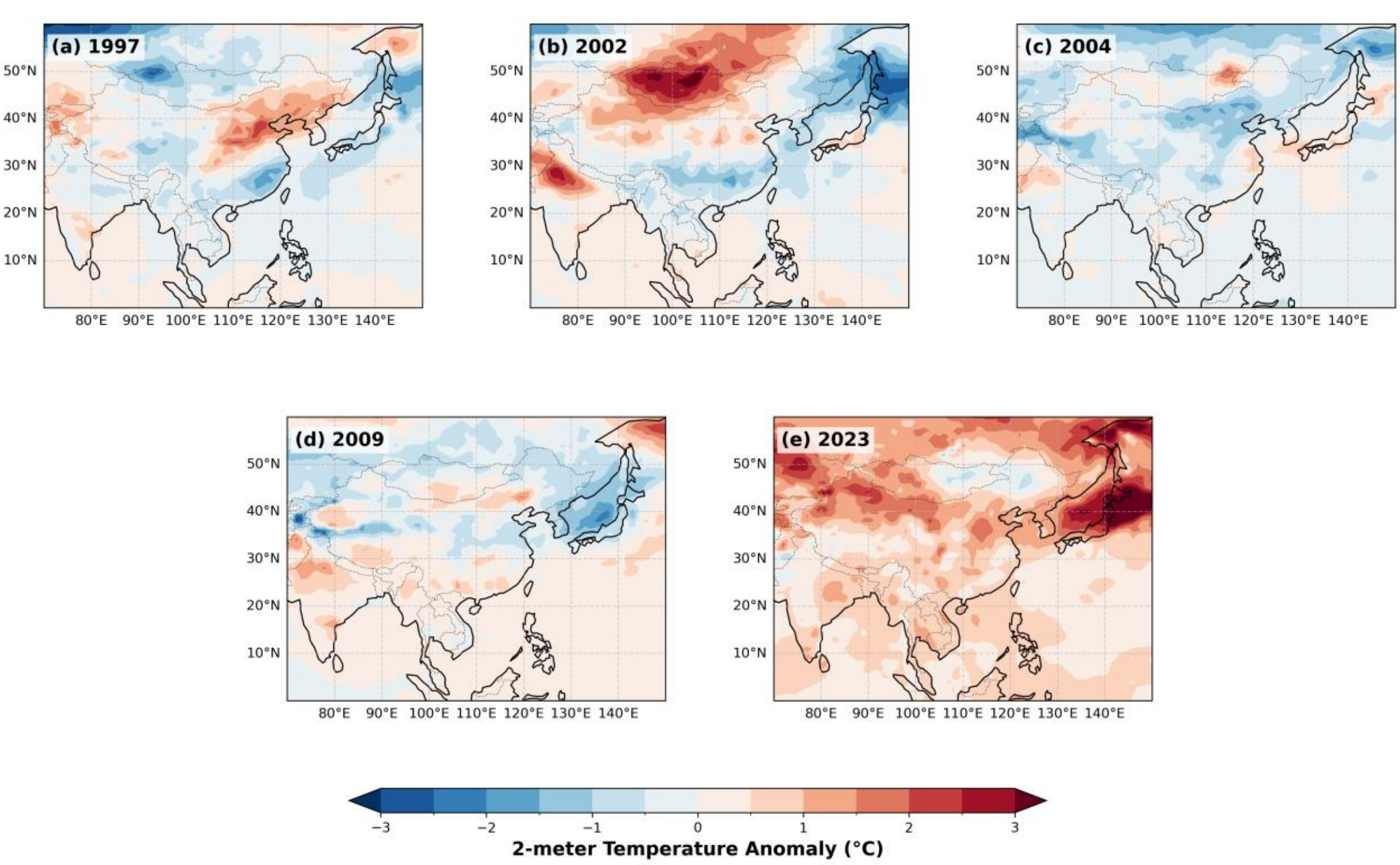

**Fig. 2.** Spatial patterns of temperature anomalies over East Asia during the developing phase of El Niño summers (July–August). The 2-meter air temperature anomalies (in °C) over East Asia averaged over July and August for the years 1997 (a), 2002 (b), 2004 (c), 2009 (d), and 2023 (e). Red (blue) shading denotes positive (negative) temperature anomalies relative to the climatological mean. For El Niño development years, only cases with Niño region SST turning from negative to positive were selected.

To anticipate the credibility of real-time operational forecasts—specifically, to determine whether a current forecast falls within a "forecast skill window," previous studies have constructed key-region SST indices based on observational data. This approach effectively identifies periods of high S2S forecast skill, recognized as forecast windows. Enhanced predictability of East Asian summer temperature can be indicated by the synergistic forcing of sea surface temperature anomalies (SSTAs) across three key oceanic regions: the tropical western Pacific, the Japan Sea-Kuroshio-Kuroshio Extension (K-KE), and the North Atlantic. When SSTAs across these three regions are concurrently positive, S2S forecast skill for East Asian summer temperatures is substantially enhanced. Physically, this positive tri-oceanic configuration corresponds to pronounced positive temperature anomalies over the central and eastern Jianghuai River Basin in China, with anomalous heat extending into Northeast Asia, encompassing the Korean Peninsula and Japan (as typified by the summer of 1994). Conversely, when SSTAs in these three regions are concurrently negative, widespread cold anomalies emerge across East Asia, as exemplified by the summer of 1993.

Based on recent observational data from June 1 to June 21 (as illustrated in Fig. 3), the precursor signals of detrended SSTAs across the three key oceanic regions are dynamically taking shape. Compared to the historical analogue year of 1994, the tropical western Pacific, the Sea of Japan-K-KE, and the North Atlantic already exhibit and maintain sufficiently pronounced anomalous warming. This suggests that East Asia's summer temperature forecast skill will stay at a relatively high level in the coming summer, thereby ensuring a high degree of credibility for real-time operational forecasts and decision-making. Furthermore, drawing upon a comparative analysis of historical analogues, the following summer is expected to feature pronounced positive temperature anomalies over central and eastern China, the Korean Peninsula, and Japan.

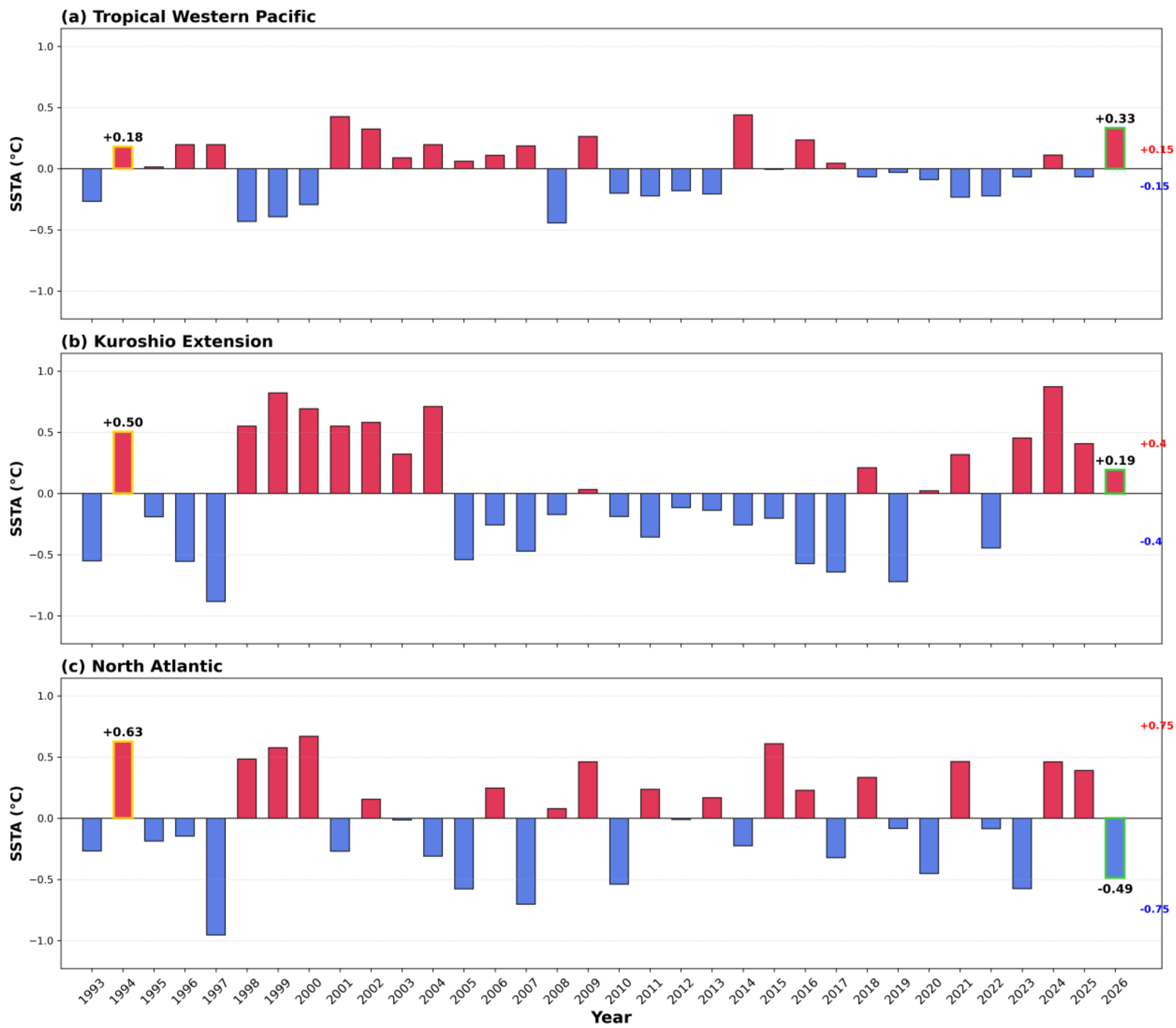

**Fig. 3.** Detrended SSTA indices across three key oceanic regions from June 1 to June 21 (1993–2026). Interannual variations of the detrended SSTA indices for (a) the Tropical Western Pacific, (b) the Kuroshio Extension, and (c) the North Atlantic. The red dashed lines denote the minimum SSTA thresholds required for the emergence of a high forecast skill window in each respective region. The historical analogue year 1994 and the current year 2026 are highlighted with gold and green outlines, respectively.

As illustrated in Fig. 3, alongside 1994 and 2026, the precursor SSTAs from June 1 to June 4 also exhibited concurrent positive signals in the western Pacific and the Sea of Japan–Kuroshio Extension during 2001, 2002, 2004, 2009, and 2024. A subsequent analysis of the North Atlantic SSTAs during the following July and August across these six years reveals a crucial prerequisite: whether positive SST anomalies in the North Atlantic can be maintained during this summer period serves as the fundamental determinant for models to achieve high prediction skill (Fig. 4, Fig.5). In Fig. 4, pronounced positive SSTAs were maintained in the North Atlantic during July and August of 1994, 2009, and 2024. Correspondingly, Fig. 5 reveals that the mean forecast skill, measured by the PCC, consistently exceeded 0.3 across these three years, indicating prolonged windows of high forecast skill.

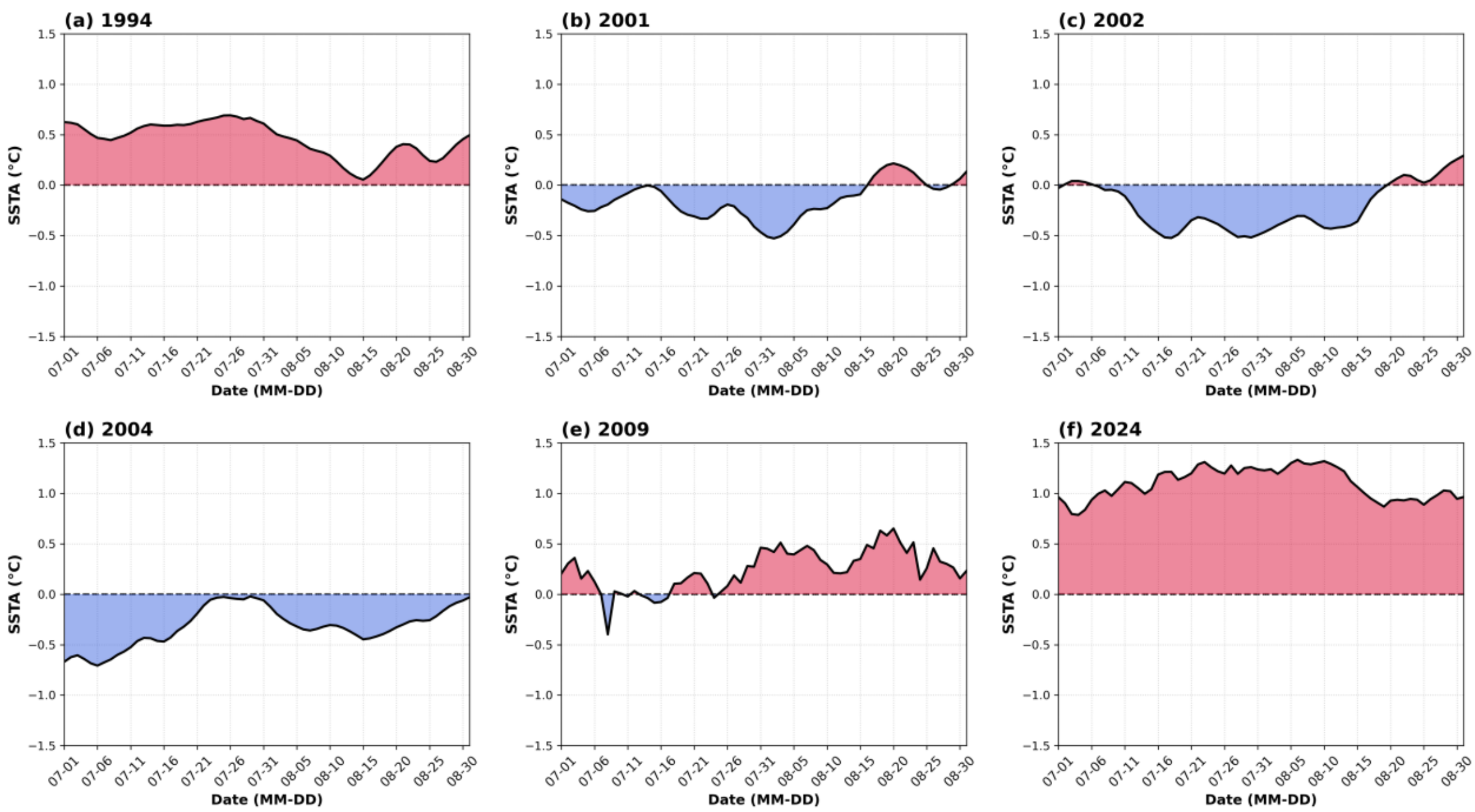

**Fig. 4.** Daily evolution of North Atlantic SSTAs during July and August across different years. Time series of the regional mean daily SSTAs (in °C) averaged over the North Atlantic from July 1 to August 31 for the years (a) 1994, (b) 2001, (c) 2002, (d) 2004, (e) 2009, and (f) 2024. Red (blue) shading denotes positive (negative) SSTA.

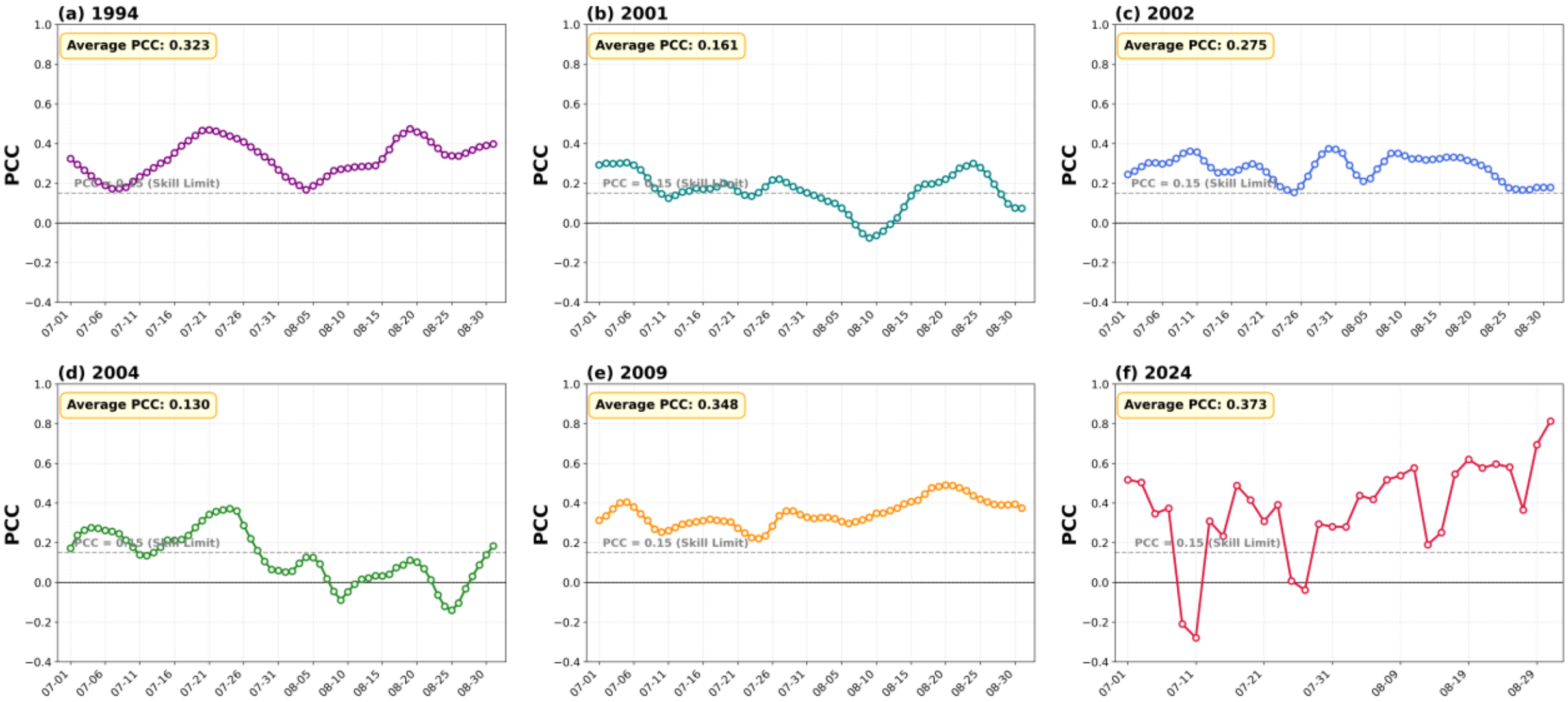

**Fig. 5.** Time series of forecast skill (PCC) for July and August forecast cases across different years. The PCC for individual forecast cases spanning July 1 to August 31 for the years (a) 1994, (b) 2001, (c) 2002, (d) 2004, (e) 2009, and (f) 2024. The PCC values in panels (a–e) are derived from the multi-model ensemble mean, whereas the results in panel f are based on the 2025 version of the ECMWF model. The value within the text box in each panel indicates the average PCC over the entire July–August period. The horizontal grey dashed line denotes the minimum threshold for effective forecast skill (PCC = 0.15). Note that the mean forecast skill (PCC) for July and August over the 30 years is approximately 0.23.

The Fig.6 displays the real-time week-3 forecast results from the EC model. Based on the results of recent forecast cases, the EC model consistently indicates upcoming high-temperature characteristics over Central and Eastern China and Japan. Analyzing the detrended SSTA precursor signals, although the North Atlantic region is currently exhibiting negative SST anomalies, it is noteworthy that the intensity of these anomalies is gradually weakening. This attenuation in anomaly intensity suggests a gradual warming trend in the North Atlantic. In conjunction with climate model predictions, it is still anticipated that warm SST anomalies will emerge in the North Atlantic during July and August. Therefore, the EC model's forecast regarding the high-temperature characteristics in Central and Eastern China and Japan is highly credible.

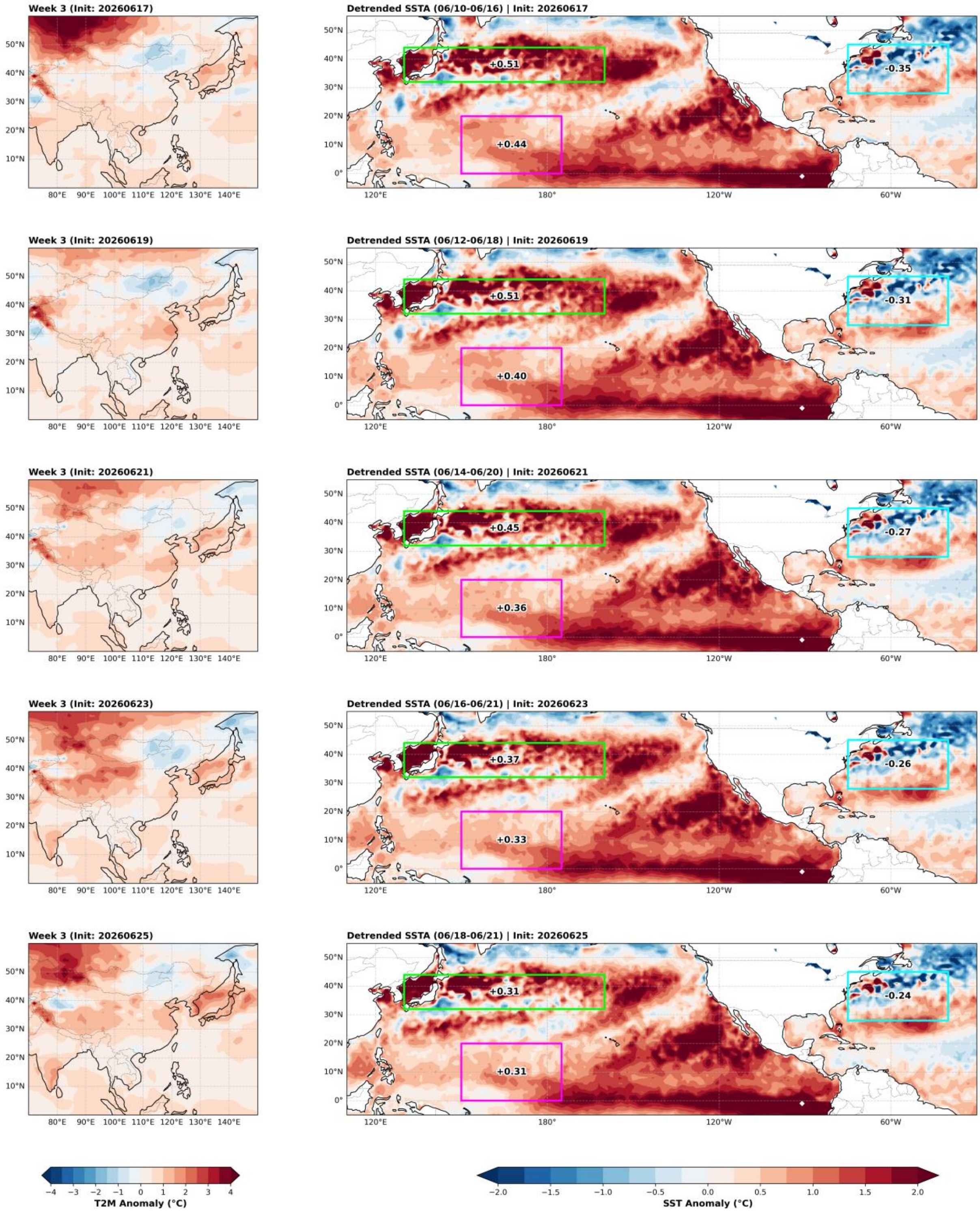

**Fig. 6.** Real-time evolution of the EC model week-3 forecasts initialized in mid-to-late June 2026. The left column displays the horizontal distribution of the T2m anomaly forecasts (°C). The right column shows the horizontal distribution of the observed detrended SSTA precursor signals (°C) for the corresponding initialization cases. The green, blue, and pink boxes represent the Sea of Japan–Kuroshio Extension region, the North Atlantic region, and the Western Pacific region, respectively, with the regional-averaged SSTA values annotated within the boxes.

In conclusion, current observational data and model predictions collectively indicate that SSTAs in these three key oceanic regions will remain positive, further corroborating the credibility of our 2026 forecast. We can deduce that forecast skill will remain at a relatively high level for the summer of 2026, thereby conferring high confidence on decision-makers who leverage model predictions. Driven by these predictions, the following summer is expected to feature pronounced positive temperature anomalies over central and eastern China, the Korean Peninsula, and Japan, which may induce regional droughts and place severe strain on power supply networks.

# 4. Conclusions and Discussion

While widespread information suggests the occurrence of a super El Niño in 2026, East Asian climate predictions are notably divergent. Accurately forecasting the evolution of summer temperatures over East Asia remains a formidable challenge. In particular, forecasts on the S2S timescale have long been constrained by the intrinsic chaotic nature of the atmosphere. To identify an effective pathway to overcome this bottleneck, this study transcends the conventional limitation of over-relying on single precursor signals, delving instead into the underlying physical mechanisms that govern the forecast skill of model predictions.

Although the latest climate monitoring and multi-model systems (including NMME and C3S) indicate the development of a pronounced El Niño event in the central and eastern equatorial Pacific during the summer of 2026, the East Asian monsoon region is situated within a highly complex global teleconnection network. Consequently, its climate anomalies are synergistically modulated by multiple factors, including tropical oceans, atmospheric circulations in the mid to high latitudes, and land surface processes. Statistical analyses of historical observational data robustly corroborate this complexity. During past summers characterized by a developing El Niño (such as 1997, 2002, and 2004), the spatial patterns of temperature anomalies across East Asia exhibited substantial inter-case variability and inconsistency. This clearly demonstrates that relying solely on El Niño as a singular precursor to anticipate East Asian summer temperatures inevitably introduces immense predictive uncertainty. Motivated by this, our study incorporates the crucial mechanism for high forecast skill windows identified in previous research. This mechanism emphasizes the synergistic forcing of SSTAs across three key oceanic regions: the tropical western Pacific, the Sea of Japan–Kuroshio Extension, and the North Atlantic. When positive SST anomalies occur concurrently across these three domains, this specific oceanic configuration effectively excites and sustains a stable EAP teleconnection pattern. Coupled with a Eurasian wave train associated with the NAO, this synergistic forcing provides a robust, persistent, and reliable source of predictability for S2S forecasts over East Asia.

To strengthen decision-makers' resolve and confidence to act, an advanced assessment of real-time forecast credibility was conducted for the summer of 2026. Such assessment draws on real-time observations of large-scale background signals. A meticulous analysis of historical analogues with similar precursor signals (such as 1994, 2009, and 2024) elucidates a crucial dynamic prerequisite: building upon preexisting warming, maintaining positive SSTAs across the North Atlantic during the subsequent July and August is the fundamental determinant for meteorological models to achieve prolonged high prediction skill. Current observational data and model predictions collectively affirm that the SSTAs in these three regions maintain positive anomalies as anticipated.

While the future evolution of SSTAs can occasionally be modulated by complex local air-sea interactions and internal dynamical variability, the robust precursor signals currently observed affirm that the specific tri-oceanic SSTA configuration capable of driving high summer prediction skill over East Asia is firmly maintained. This persistent positive state suggests that the East Asia summer temperature forecast skill will stay at a relatively high level in the coming summer. Admittedly, S2S forecast skill is intrinsically constrained; thus, the relatively high skill referenced here only indicates a PCC above 0.15, which remains well below the level of a perfect forecast. Nevertheless, this already represents the current upper limit of S2S predictive capability.

The maintenance of this configuration confers a high degree of credibility upon the operational forecasts, predicting that the following summer is expected to feature pronounced positive temperature anomalies over central and eastern China, the Korean Peninsula, and Japan. Consequently, this finding not only provides a robust physical basis for predicting the anomalous warm summer but also profoundly demonstrates that advance assessment of real-time forecast credibility can potentially strengthen decision-makers' resolve to act decisively against impending regional droughts and severe strain on power supply networks. This ultimately establishes a highly valuable scientific paradigm and predictive pathway for significantly enhancing both climate prediction skill and its practical usability on the S2S timescale.